\documentclass[aps,prd,nofootinbib]{revtex4}

\usepackage{graphicx}
\usepackage{slashed}

\begin{document}

\title{Bottomonium production associated with a photon at a high luminosity $e^+e^-$ collider with one-loop QCD correction}

\author{Zhan Sun}
\author{Xing-Gang Wu}
\email{wuxg@cqu.edu.cn}
\author{Gu Chen}
\author{Yang Ma}
\author{Hong-Hao Ma}
\author{Huan-Yu Bi}

\address{Department of Physics, Chongqing University, Chongqing 401331, P.R. China}

\date{\today}

\begin{abstract}

We make a detailed discussion on the one-loop QCD correction to the bottomonium production associated with a photon, i.e. via the channel $e^{+}e^{-} \to\gamma^*/Z^0 \to |H_{b\bar{b}}\rangle + \gamma$, where $|H_{b\bar{b}}\rangle$ stands for the color-singlet bottomonium state as $\eta_b$, $\Upsilon$, $h_b$ or $\chi_{bJ}$ ($J$=0, 1 or 2), respectively. At the super $Z$ factory with the collision energy $E_{cm} \sim m_Z$, by summing up the cross sections for all bottomonium states, we obtain a large one-loop QCD correction, i.e. $|R|\sim 30\%$. This ensures the necessity and importance of the one-loop QCD corrections for the present processes. Further more, for the $\eta_b$, $h_b$ and $\chi_{bJ}$ production, their cross sections are dominated by the $s$-channel diagrams and are enhanced by the $Z^0$ boson resonance effect when $E_{cm}\sim m_Z$. While, for the $\Upsilon$ production, such resonance effect shall be smeared by a large $t(u)$-channel contribution that dominant over the $s$-channel one. Theoretical uncertainties caused by slight change of $E_{cm}$, the $b$-quark mass, the renormalization scale and etc. have been presented. At the super $Z$ factory with a high luminosity up to ${\cal L}=10^{36}{\rm cm}^{-2}{\rm s}^{-1}$, the bottomonium plus one photon events are sizable, especially for $\eta_b$ and $\Upsilon$, which have large signal significance. Summing up all bottomonium states, we shall totally have $\sim 3.8\times10^{5}$ bottomonium events in one operation year. So, the super $Z$ factory shall provide a good platform for studying the bottomonium properties. \\


\end{abstract}

\maketitle

\section{Introduction}

Within the framework of the non-relativistic QCD factorization theory (NRQCD), the heavy quarkonium production or decay can be factorized into a sum of products of short-distance coefficients and the non-perturbative but universal long-distance matrix elements~\cite{NRQCD}. The short-distance coefficients are perturbatively calculable, which can be expanded in a combined power series of the strong coupling constant ($\alpha_s$) and the relative velocity of the constituent quarks in the quarkonium rest frame $(v)$. The top quark is too heavy to form a steady bound state before it decays, thus, the bottomonium is the heaviest and most compact bound state of quark-antiquark system within the standard model. Both its typical coupling constant and relative velocity are smaller than those of charmonium. In general, the perturbative results for bottomonium will be more convergent over the expansion of $\alpha_s$ and $v^2$ than the charmonium cases. Thus, if at an experimental platform enough bottomonium events could be generated, it shall provide a relatively more confidential test of NRQCD.

At the $B$ factory, the bottomonium production via the channel $e^+e^- \to \gamma^* \to |H_{b\bar{b}}\rangle+\gamma$ with one-loop QCD correction has been studied~\cite{ccY3,ccY4,ccY5}, where $|H_{b\bar{b}}\rangle$ equals to $\eta_b$ or $\chi_{bJ}$ $(J=1,2,3)$ that has the charge-conjugation parity $C=+1$. Because the bottomonium mass is close to the threshold of the $B$ factories as Belle and BABAR, the emitted photons could be soft (this is caused by the $2\to 2$ phase-space constraint). Thus, the pQCD estimations on the bottomonium production at the $B$ factories may be questionable and some extra treatments have to be introduced in order to achieve a reliable pQCD estimation. For example, in Ref.\cite{ccY3}, an extra artificial phase-space factor has been suggested to suppress the singular contributions from the phase-space end-point region, e.g.  $\left(1-(M^2_{|H_{b\bar{b}} \rangle}/s)^2 \right)^3/ \left(1-(4m^2_b/s)^2\right)^3$ for $\eta_b$ production and $\left(1-(M^2_{|H_{b\bar{b}} \rangle}/s)^2 \right)/ \left(1-(4m^2_b/s)^2\right)$ for $\chi_{bJ}$ production have been introduced accordingly. On the other hand, the super $Z$ factory, which runs at a much higher collision energy and with a high luminosity up to $10^{34-36}{\rm cm}^{-2}{\rm s}^{-1}$~\cite{zfactory}, provides a more reliable platform for studying the bottomonium properties. The bottomonium production via semi-exclusive channels as $e^{+}e^{-} \rightarrow |H_{b\bar{b}}\rangle +X$ with $X=b+\bar{b}$ or $g+g$, has shown one of such examples~\cite{inclucivecharmonium}. Moreover, in Refs.\cite{ccY1,ccY2}, the authors concluded that a large amount of charmonium events can be produced via the charmonium plus photon channel at the super $Z$ factory. It is natural to assume that we shall also observe enough bottomonium events at this platform. One can treat the bottomonium in a similar way as that of charmonium, however, it has its own properties and specific points, deserving a separate discussion.

For the purpose, we shall calculate the production channel $e^+e^- \to \gamma^*/Z^0 \to |H_{b\bar{b}}\rangle + \gamma$ with one-loop QCD correction. Different to previous treatment~\cite{ccY3,ccY4,ccY5}, where only the $s$-channel diagrams have been calculated, we shall discuss both the $s$-channel and $t(u)$-channel diagrams, including their cross terms, so as to achieve a sound estimation. We shall show that the $t(u)$-channel diagrams (or roughly, the initial state radiation diagrams~\cite{ccY6}) can also provide sizable contributions. Due to angular momentum conservation and Bose statistics (known as the Landau-Pomeranchuk-Yang theorem~\cite{lpy}), those channels via $\gamma^*$ propagator are forbidden, i.e. the $s$-channels of $e^+e^- \to\gamma^* \to \Upsilon\;(h_b) + \gamma$ and the $t(u)$-channels of $e^+e^- \to \gamma^* \to \eta_b \; (h_b,\chi_{bJ}) + \gamma$. The color-octet contributions shall be greatly suppressed in comparison to the color-singlet cases, which mainly come from the suppression of the color-octet matrix element to the color-singlet matrix element, i.e. about $v_b^4$ suppression to the corresponding color-singlet matrix element~\cite{NRQCD} with $v_b \sim 0.1$ being the relative velocity between the constituent quarks in bottomonium. In addition, at the present one-loop level, only real corrections are non-zero for the color-octet channels, so there are also extra phase-space and color- suppression to the color-singlet cases. In the present paper, we shall not take the color-octet quarkonium states into consideration.

The remaining parts of the paper are organized as follows. In Sec.II, we present the calculation technology for dealing with the bottomonium plus one photon production at the tree-level and the one-loop level, respectively. In Sec.III, we present our numerical results. Sec.IV is reserved for a summary.

\section {calculation technology}

The cross section for $e^+(p_{2})+e^-(p_{1})\to  |H_{b\bar{b}}\rangle(p_{3})+\gamma(p_{4})$ can be factorized as~\cite{NRQCD,NRQCD2}
\begin{equation}
d\sigma=\sum_{n} d\hat\sigma(e^+e^- \to (b\bar{b})[n]+\gamma) \langle{\cal O}^H(n) \rangle \;,
\end{equation}
where $\langle{\cal O}^H(n) \rangle$ stands for the probability of the perturbative state $(b\bar{b})[n]$ into the bound state $|H_{b\bar{b}}\rangle$ with the same quantum number $[n]$. The matrix elements are non-perturbative and universal, which can be determined by comparing with the experimental data. Theoretically, the color-singlet ones can be related to the wavefunction at the origin for $S$-wave states or the first derivative of the wavefunction at the origin for $P$-wave states, which can be estimated via phenomenological potential model.

The differential cross section $d\hat\sigma(e^+e^- \to (b\bar{b})[n]+\gamma)$ stands for the $2\to2$ short-distance cross section, i.e.
\begin{eqnarray}
&& d\hat\sigma(e^{+}+e^{-}\rightarrow (b\bar{b})[n]+\gamma) = \frac{1}{4\sqrt{(p_1\cdot p_2)^2-m_{e^+}^2 m_{e^-}^2}} \overline{\sum}  |{\cal M}|^{2} d\Phi_2 \;,
\end{eqnarray}
where $d\Phi_2$ refers to the two-body phase space and the symbol $\overline{\sum}$ means we need to average over the initial degrees of freedom and sum over the final ones.

\subsection{The tree-level calculation}

\begin{figure*}
\includegraphics[width=0.9\textwidth]{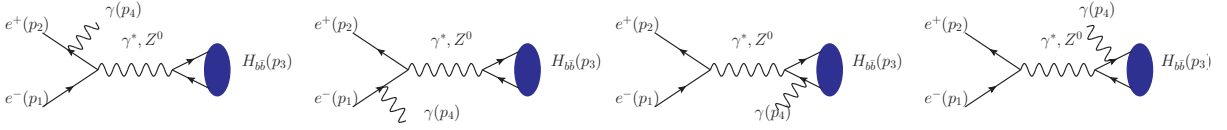}
\caption{Tree-level Feynman diagrams for $e^+e^- \to\gamma^*/Z^0 \to |H_{b\bar{b}} \rangle + \gamma$, where $|H_{b\bar{b}}\rangle$ denotes for $\eta_b$, $\Upsilon$, $h_b$ and $\chi_{bJ}$ ($J$=0,1,2), respectively. The left two diagrams are $t(u)$-channel, and the leaving ones are $s$-channel. } \label{bbYLO}
\end{figure*}

As shown in Fig.(\ref{bbYLO}), at the tree level, there are four Feynman diagrams for $e^+e^- \to |H_{b\bar{b}}\rangle + \gamma$, where $|H_{b\bar{b}}\rangle$ denotes for $\eta_b$, $\Upsilon$, $h_b$ and $\chi_{bJ}$ ($J$=0,1,2), respectively. Two are $s$-channel diagrams and two are $t(u)$-channel ones. For those diagrams, the propagator in between can be either a virtual photon or a $Z^0$ boson. Since the behaviors of the hard scattering matrix elements for the $s$-channel and $t(u)$-channel are quite different, we shall treat them separately.

Firstly, the hard scattering matrix elements for the $s$-channel diagrams can be written as
\begin{equation}
i{\cal M}={\cal C} \; L_{rr^{\prime}}^{\mu}D_{\mu\nu}\sum^{j_{max}}_{j=1} {\cal A}_j^{\nu} , \label{MM}
\end{equation}
where ${\cal C}$ is an overall factor and the leptonic current is,
\begin{equation}
L_{rr^{\prime}}^{\mu} = \bar{v}_r(p_2) \Gamma^{\mu} u_{r^{\prime}}(p_1) ,
\end{equation}
where the indices $r$ and $r^{\prime}$ stand for the spin projections of the initial electron and positron. $j_{max}$ equals to $2$ for the tree-level, and it equals to $8$ for the one-loop QCD correction. For the production via the $Z^0$ propagator, we have $\Gamma^{\mu}= \gamma^\mu(1-4\sin^2\theta_w-\gamma^5)$ and $D_{\mu\nu} =-{i}\left(g_{\mu\nu}-{k_\mu k_\nu} /{k^2}\right)/\left(k^2-m^2_Z +im_Z\Gamma_z\right)$, where $\Gamma_z$ stands for the total decay width of $Z^0$ boson. While, for the production via the virtual photon propagator, we have $\Gamma^{\mu}=\gamma^\mu$ and $D_{\mu\nu}=-{i g_{\mu\nu}}/{k^2}$.

For specific bottomonium production, the expressions for the $s$-channel ${\cal A}_j^{\nu}$ are
\begin{widetext}
\begin{eqnarray}
{\cal A}_1^{\nu (S = 0,L = 0)}(\eta_b) &=& i\;\textrm{Tr} \left.\left[ {\Pi _{{p_3}}^0(q){\Gamma^\nu_{b\bar{b}} }\frac{{-\slashed{p}_{32}-\slashed{p}_4} + {m_b}}
{{(p_{32}+p_4)^2 - m_b^2}}\not\!\epsilon ({p_4})} \right]\right|_{q = 0}\;, \\
{\cal A}_2^{\nu (S = 0,L = 0)}(\eta_b) &=& i\;\textrm{Tr} \left.\left[ {\Pi _{{p_3}}^0(q)\not\!\epsilon ({p_4})\frac{{\slashed{p}_{31}+\slashed{p}_4} + {m_b}}{{(p_{31}+p_4)^2 - m_b^2}}{\Gamma^\nu_{b\bar{b}} }}\right]\right|_{q = 0} \;, \\
{\cal A}_1^{\nu (S = 1,L = 0)}(\Upsilon) &=& i\; \varepsilon_{s,\beta}(p_3) \textrm{Tr} \left.\left[{\Pi _{{p_3}}^{\beta}(q){\Gamma^\nu_{b\bar{b}} }\frac{{-\slashed{p}_{32}-\slashed{p}_4} + {m_b}}
{{(p_{32}+p_4)^2 - m_b^2}}\not\!\epsilon ({p_4})} \right]\right|_{q = 0}\;, \\
{\cal A}_2^{\nu (S = 1,L = 0)}(\Upsilon) &=& i\; \varepsilon_{s,\beta}(p_3) \textrm{Tr} \left.\left[{\Pi _{{p_3}}^{\beta}(q)\not\!\epsilon ({p_4})\frac{{\slashed{p}_{31}+\slashed{p}_4} + {m_b}}{{(p_{31}+p_4)^2 - m_b^2}}{\Gamma^\nu_{b\bar{b}} }} \right]\right|_{q = 0} \;,
\end{eqnarray}
\end{widetext}
and
\begin{widetext}
\begin{eqnarray}
{\cal A}_1^{\nu (S = 0,L = 1)}(h_b) &=& i\;\varepsilon_{l,\alpha}(p_3) \frac{d}{dq_{\alpha}} \textrm{Tr}\left.\left[{\Pi _{{p_3}}^0(q){\Gamma^\nu_{b\bar{b}} }\frac{{-\slashed{p}_{32}-\slashed{p}_4} + {m_b}} {{(p_{32}+p_4)^2 - m_b^2}}\not\!\epsilon ({p_4})} \right]\right|_{q = 0}\;, \\
{\cal A}_2^{\nu (S = 0,L = 1)}(h_b) &=& i\;\varepsilon_{l,\alpha}(p_3) \frac{d}{dq_{\alpha}} \textrm{Tr} \left.\left[{\Pi _{{p_3}}^0(q) \not\!\epsilon ({p_4})\frac{{\slashed{p}_{31}+\slashed{p}_4} + {m_b}}{{(p_{31}+p_4)^2 - m_b^2}}{\Gamma^\nu_{b\bar{b}} }} \right] \right|_{q = 0}\;, \\
{\cal A}_1^{\nu (S = 1,L = 1)}(\chi_{bJ}) &=& i\;\varepsilon^{J}_{\alpha\beta}(p_3) \frac{d}{dq_{\alpha}} \textrm{Tr}\left.\left[{\Pi _{{p_3}}^{\beta}(q){\Gamma^\nu_{b\bar{b}} }\frac{{-\slashed{p}_{32}-\slashed{p}_4} + {m_b}} {{(p_{32}+p_4)^2 - m_b^2}}\not\!\epsilon ({p_4})} \right]\right|_{q = 0} \;,\\
{\cal A}_2^{\nu (S = 1,L = 1)}(\chi_{bJ}) &=& i\;\varepsilon^{J}_{\alpha\beta}(p_3)\frac{d}{dq_{\alpha}} \textrm{Tr}\left.\left[{\Pi _{{p_3}}^{\beta}(q) \not\!\epsilon({p_4})\frac{{\slashed{p}_{31}+ \slashed{p}_4} + {m_b}}{{(p_{31}+p_4)^2 - m_b^2}}{\Gamma^\nu_{b\bar{b}} }} \right]\right|_{q = 0}\;.
\end{eqnarray}
\end{widetext}
The interaction vertex
\begin{equation}
\Gamma^{\nu}_{b\bar{b}}=\gamma^{\nu}(\xi_1 P_L+\xi_2 P_R) ,
\end{equation}
where $P_L={(1-\gamma^5)}/{2}$ and $P_R={(1+\gamma^5)}/{2}$. Here $\xi_1=2-\frac{4}{3} \sin^2\theta_w$ and $\xi_2=-\frac{4}{3}\sin^2\theta_w$ for $(Z{b}{\bar{b}})$-vertex, $\xi_1=1$ and $\xi_2=1$ for $(\gamma^{*}{b}{\bar{b}})$-vertex. The momenta of the constituent quarks are
\begin{equation}
p_{31} = \frac{m_b}{M_{b\bar{b}}}{p_3} + q \;\;{\rm and}\;\;
p_{32} = \frac{m_{\bar b}}{M_{b\bar{b}}}{p_3} - q,
\end{equation}
where the bottomonium mass $M_{b\bar{b}}= m_b + m_{\bar{b}}$, $q$ is the relative momentum between the two constituent quarks inside the bottomonium.

The covariant form of the projectors are~\cite{spinprojector1,spinprojector2},
\begin{displaymath}
\Pi _{{p_3}}^0(q)=\frac{1} {{\sqrt {8m_b^3} }} \left(\frac{{{{\not\!p}_3}}}
{2} - {\not\!q } - {m_b}\right){\gamma ^5} \left(\frac{{{{\not\!p}_3}}}
{2} + {\not\!q } + {m_b}\right)
\end{displaymath}
and
\begin{displaymath}
\Pi _{{p_3}}^{\beta}(q)=\frac{1} {{\sqrt {8m_b^3} }} \left(\frac{{{{\not\!p}_3}}} {2} - {\not\!q } - {m_b}\right) {\gamma ^\beta } \left(\frac{{{{\not\!p}_3}}} {2} + {\not\!q } + {m_b}\right)\;.
\end{displaymath}
The sum over the polarization for a spin-triplet $S$-wave state or a spin-singlet $P$-wave state is given by,
\begin{equation}
\Pi_{\alpha\beta} =\sum_{J_z}\varepsilon_\alpha \varepsilon^*_{\beta} =-g_{\alpha\beta}+\frac{p_{3\alpha} p_{3\beta}}{p_{3}^2}\;,
\end{equation}
where $\varepsilon$ stands for the polarization vector $\varepsilon_l$ or $\varepsilon_s$ respectively. The sum over the polarization for the spin-triplet $P$-wave states ($^3P_J$ with $J=0,1,2$) is given by~\cite{NRQCD2,spinprojector4},
\begin{eqnarray}
\varepsilon^{(0)}_{\alpha\beta} \varepsilon^{(0)*}_{\alpha'\beta'} &=& \frac{1}{3} \Pi_{\alpha\beta}\Pi_{\alpha'\beta'}, \\
\sum_{J_z}\varepsilon^{(1)}_{\alpha\beta} \varepsilon^{(1)*}_{\alpha'\beta'} &=& \frac{1}{2}(\Pi_{\alpha\alpha'}\Pi_{\beta\beta'}- \Pi_{\alpha\beta'}\Pi_{\alpha'\beta}), \\
\sum_{J_z}\varepsilon^{(2)}_{\alpha\beta} \varepsilon^{(2)*}_{\alpha'\beta'} &=& \frac{1}{2}({\Pi_{\alpha\alpha'}\Pi_{\beta\beta'}+ \Pi_{\alpha\beta'}\Pi_{\alpha'\beta}}) -\frac{1}{3}\Pi_{\alpha\beta}\Pi_{\alpha'\beta'}\;.
\end{eqnarray}

Secondly, the hard scattering matrix elements for the $t(u)$-channel diagrams can be written as
\begin{equation}
i{\cal M}={\cal C} \; \left[L_{rr^{\prime}}^{\mu}(1)+L_{rr^{\prime}}^{\mu}(2) \right] D_{\mu\nu}{\cal A}^{\nu} , \label{MM1}
\end{equation}
where the leptonic currents are,
\begin{equation}
L_{rr^{\prime}}^{\mu}(1) = \bar{v}_r(p_2)\not\!\epsilon ({p_4})\frac{{\slashed{p}_{4} -\slashed{p}_2} + {m_{e^-}}}{{(p_{4}-p_2)^2 - m_{e^-}^2}} \Gamma^{\mu} u_{r^{\prime}}(p_1)
\end{equation}
and
\begin{equation}
L_{rr^{\prime}}^{\mu}(2) = \bar{v}_r(p_2) \Gamma^{\mu} \frac{{\slashed{p}_{1} -\slashed{p}_4} + {m_{e^+}}}{{(p_{1}-p_4)^2 - m_{e^+}^2}}\not\!\epsilon ({p_4}) u_{r^{\prime}}(p_1).
\end{equation}
The reduced amplitude ${\cal A}^{\nu}$ for each bottomonium state takes the form
\begin{eqnarray}
&&{\cal A}^{\nu (S = 0,L = 0)}(\eta_b) = i\;\textrm{Tr} \left.\left[ \Pi _{{p_3}}^0(q){\Gamma^\nu_{b\bar{b}}} \right]\right|_{q = 0}, \\
&&{\cal A}^{\nu (S = 1,L = 0)}(\Upsilon) = i\; \varepsilon_{s,\beta}(p_3)\textrm{Tr} \left.\left[\Pi _{{p_3}}^{\beta}(q){\Gamma^\nu_{b\bar{b}}} \right]\right|_{q = 0}, \\
&&{\cal A}^{\nu (S = 0,L = 1)}(h_b) = i\;\varepsilon_{l,\alpha}(p_3)\frac{d}{dq_{\alpha}} \textrm{Tr}\left.\left[{\Pi _{{p_3}}^0(q){\Gamma^\nu_{b\bar{b}}}} \right]\right|_{q = 0}\;,  \\
&&{\cal A}^{\nu (S = 1,L = 1)}(\chi_{bJ}) = i\;\varepsilon^{J}_{\alpha\beta}(p_3) \frac{d}{dq_{\alpha}} \textrm{Tr}\left.\left[ {\Pi _{{p_3}}^{\beta}(q){\Gamma^\nu_{b\bar{b}} }} \right]\right|_{q = 0} \;.
\end{eqnarray}

As for the parameter ${\cal C}$ defined in Eqs.(\ref{MM},\ref{MM1}), we have ${\cal C}=\frac{e_b e^3} {\sin^2\theta_w\cdot (4\cos\theta_w)^2}$ and ${\cal C}={e^2_b e^3}$ for the $s$-channel via $Z^0$ propagator and via $\gamma^*$ propagator, respectively; ${\cal C}=\frac{e^3} {\sin^2\theta_w\cdot (4\cos\theta_w)^2}$ and ${\cal C}={e_b e^3}$ for the $t(u)$-channel via $Z^0$ propagator and $\gamma^*$ propagator, respectively. Here $e_b$ stands for the electric charge of $b$ quark in unit $e$.

\subsection{The one-loop QCD calculation}

\begin{figure*}[htb]
\includegraphics[width=0.90\textwidth]{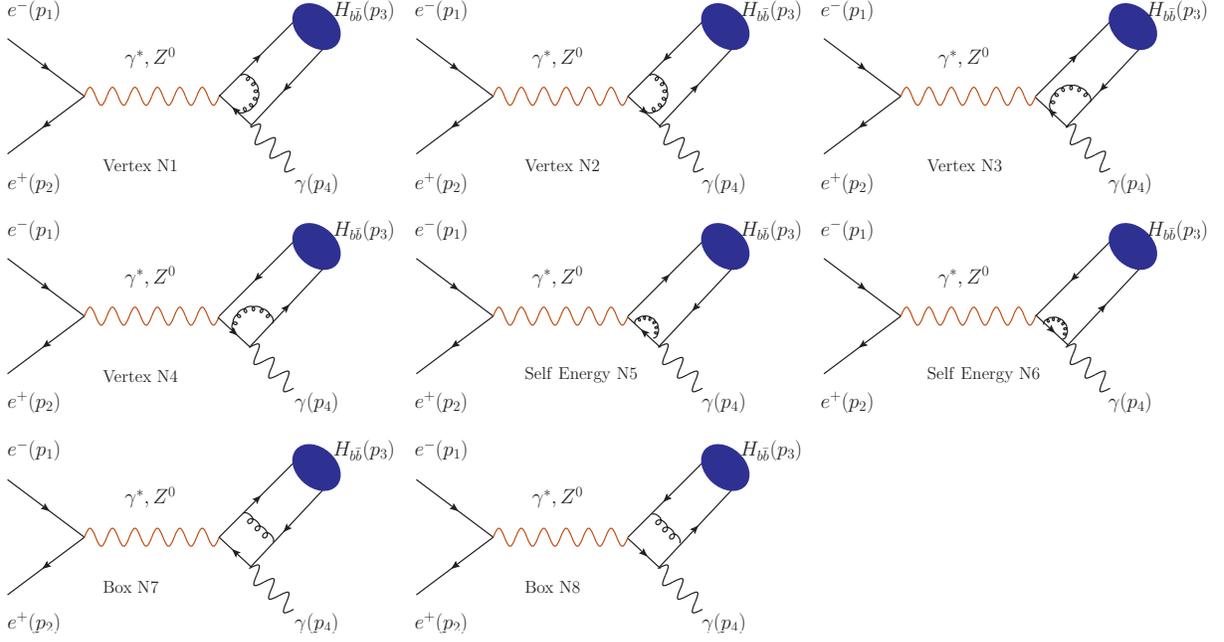}
\caption{One-loop QCD correction to the $s$-channel diagrams of $e^+e^- \to \gamma^*/Z^0\to |H_{b\bar{b}}\rangle +\gamma$, where $|H_{b\bar{b}}\rangle$ denotes $\eta_b$, $\Upsilon$, $h_b$ and $\chi_{bJ}$ $(J=0,1,2)$, respectively. } \label{bbYsNLO}
\end{figure*}

\begin{figure}[htb]
\includegraphics[width=0.60\textwidth]{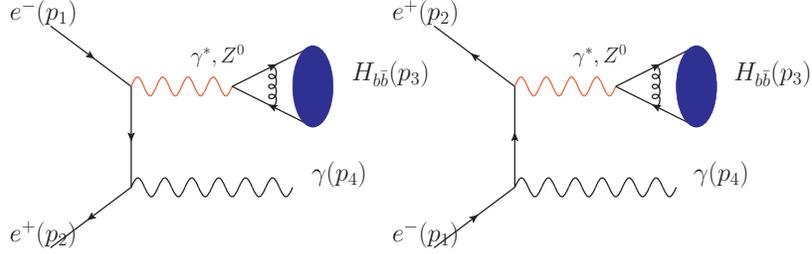}
\caption{One-loop QCD correction to the $t(u)$-channel diagrams of $e^+e^- \to \gamma^*/Z^0\to |H_{b\bar{b}}\rangle +\gamma$, where $|H_{b\bar{b}}\rangle$ denotes $\eta_b$, $\Upsilon$, $h_b$ and $\chi_{bJ}$ $(J=0,1,2)$, respectively. } \label{bbYtNLO}
\end{figure}

We proceed to calculate the one-loop QCD correction for the process $e^+e^- \to |H_{b\bar{b}}\rangle +\gamma$. It is noted that the $t(u)$-channel diagrams also have sizable contributions, especially at the super $Z$ factory. Thus, it is necessary to take both the $s$- and $t(u)$-channels into consideration so as to achieve a sound estimation.

Due to vanishing of the color factor, there is no real corrections for $e^+e^- \to |H_{b\bar{b}} \rangle + \gamma$ at the one-loop level. The cross section can be schematically written as:
\begin{eqnarray}
d\sigma=d\sigma_{Born} + d\sigma_{Virtual} + {\cal O}(\alpha_s^2)
\end{eqnarray}
with
\begin{equation}
d\sigma_{Born} = \frac{1}{2s}\overline{\sum}|{\cal M}_{Born}|^2 {\textrm{d}\Phi}_2
\end{equation}
and
\begin{equation}
d\sigma_{Virtual} = \frac{1}{s}\overline{\sum} {\cal M}_{Born}{\cal M}^*_{Virtual} \textrm{d}\Phi_2 ,
\end{equation}
where $s=(p_1+p_2)^2$. ${\cal M}_{Born}$ refers to the tree-level amplitudes and ${\cal M}_{Virtual}$ stands for the amplitudes of the virtual corrections.

The Feynman diagrams for the one-loop QCD correction to the process $e^+e^- \to |H_{b\bar{b}} \rangle +\gamma$ are shown in Figs.(\ref{bbYsNLO},\ref{bbYtNLO}). At the one-loop level, there are eight virtual diagrams for the $s$-channel and two more virtual diagrams for the $t(u)$-channel. Among these diagrams, the self-energy diagrams (N5, N6) contain ultraviolet (UV) divergence, the vertex diagrams (N1-N4 and the two $t(u)$-channel diagrams) contain both UV and IR divergences, while the box diagrams (N7, N8) have infrared (IR) divergences. Moreover, in the box diagrams (N7, N8) and the two $t(u)$-channel diagrams, the gluon exchange between the constituent quark and antiquark of bottomonium leads to Coulomb singularities, which can be absorbed into the redefinition of the bottomonium matrix elements. In our calculation, the usual dimensional renormalization scheme is adopted ($D=4-2\epsilon$), which can be further performed in on-mass-shell (OS) scheme such that the external quark lines do not receive any QCD corrections. As for the channels via the $Z^0$ boson, we shall meet the $\gamma^5$ problem, which can be treated following the idea of Refs.\cite{gamma5problem1, gamma5problem2, gamma5problem3,gamma5problem4, gamma5problem5,gamma5problem6}. The subtle points for the treatment of $\gamma_5$ have been listed in Ref.\cite{ccY2}. After canceling all those UV and IR divergences and absorbing the Coulomb singularities, we can get the finite results for the mentioned production processes.

More specifically, we show some more detail on how to deal with the Coulomb singularities. The total cross section of the process can be schematically written as
\begin{eqnarray}
\sigma &=& \langle{\cal O}^H(n) \rangle_0 \cdot \hat{\sigma}^{(0)}\left( 1+ \frac{4\pi\alpha_s}{3v} + \frac{\alpha_s \hat{C}}{\pi} \right) + {\cal O}(\alpha_s^2) \nonumber\\
&=& \langle{\cal O}^H(n) \rangle_0 \cdot \hat{\sigma}^{(0)}\left( 1+ \frac{4\pi\alpha_s}{3v}\right) \left(1 + \frac{\alpha_s \hat{C}}{\pi} \right)  + {\cal O}(\alpha_s^2),
\end{eqnarray}
where $\langle{\cal O}^H(n) \rangle_0$ is the tree-level non-perturbative but universal matrix element which represents the hadronization probability of the perturbative state $(b\bar{b})[n]$ into the bound state $H$. $\hat{\sigma}^{(0)}$ is the hard part of the tree-level cross section. It is noted that the overall Coulomb correction to the tree-level matrix element is~\cite{NRQCD2}
\begin{equation}
\langle{\cal O}^H(n) \rangle_{R}=\left( 1+ \frac{4\pi\alpha_s}{3v}\right)\langle{\cal O}^H(n) \rangle_0 .
\end{equation}
In combination with the above two equations, the singular $1/v$ terms exactly cancel, i.e.
\begin{eqnarray}
\sigma &=& \left[\langle{\cal O}^H(n) \rangle_{R} \left( 1- \frac{4\pi\alpha_s}{3v}\right)\right] \left( 1+ \frac{4\pi\alpha_s}{3v}\right) \left(1 + \frac{\alpha_s \hat{C}}{\pi} \right) \cdot \hat{\sigma}^{(0)} + {\cal O}(\alpha_s^2) \nonumber\\
&=& \langle{\cal O}^H(n) \rangle_{R} \cdot \hat{\sigma}^{(0)} \left( 1 + \frac{\alpha_s \hat{C}}{\pi} \right) + {\cal O}(\alpha_s^2) ,
\end{eqnarray}
The color-singlet matrix elements can be related with the wavefunction at the origin for $S$-wave bottomonium or the first derivative of the wavefunction at the origin for $P$-wave bottomonium~\cite{NRQCD,NRQCD2},
\begin{displaymath}
\frac{\langle 0|{\cal O}_{\bf 1}^{\eta_b}|0 \rangle_{R}}{2N_c} \simeq \frac{\langle 0|{\cal O}_{\bf 1}^{\Upsilon}|0 \rangle_{R}}{6N_c} = \frac{|R_{1S}(0)|^2}{4\pi}
\end{displaymath}
and
\begin{displaymath}
\frac{\langle 0|{\cal O}_{\bf 1}^{\chi_{bJ}}|0 \rangle_{R}}{2(2J+1) N_c} \simeq \frac{\langle 0|{\cal O}_{\bf 1}^{h_{b}}|0 \rangle_{R}}{6N_c} =\frac{3|R'_{1P}(0)|^2}{4\pi} ,
\end{displaymath}
where $N_c=3$ and $J=0,1,2$, respectively.

All calculations are done automatically via proper using of Fortran or Mathematica packages. The loop calculation technologies adopted here have been described in detail in Ref.\cite{ccY2}. For self-consistency, we present the main points here. The interesting reader may turn to Ref.\cite{ccY2} for detailed technology. More explicitly, we adopt \textbf{FeynArts}~\cite{feynarts1,feynarts2} to generate Feynman diagrams and the analytical amplitudes with or without one-loop QCD correction. Then the package \textbf{FeynCalc}~\cite{feyncalc} is applied to simplify the trace of the $\gamma$-matrixes for the close fermion quark-line loop such that the hard scattering amplitudes are transformed into expressions with the tensor integrals over the loop momentum $l$. By means of the Mathematica function \textbf{TIDL}~\cite{feyncalc}, those tensor integrals can be further reduced to expressions of the scalar products of all independent momenta, such as $l \cdot p_3, l \cdot p_4$ and $l^2$. Finally, with the help of the Mathematica package \textbf{\$}\textbf{Apart}~\cite{Apart,Apart2} together with the Feynman integral reduction algorithm \textbf{FIRE}~\cite{fire1,fire2}, those complicated scalar integrals can be reduced to the simple/conventional irreducible master integrals~\cite{dimensionalrenormalization}. For convenience, we put the necessary analytical expressions of all the master integrals in Appendix A.

\section{Numerical results}

In doing numerical calculation, the $b$-quark mass is taken as, $m_b=4.90^{+0.10}_{-0.10}$ GeV. This choice of $m_b$ is consistent with the suggestion of the so-called $1S$-mass~\cite{bquarkmass1} and the potential-model estimations~\cite{bquarkmass2,bquarkmass3}. The bottomonium wavefunctions at the origin (for $S$-wave states) and their first derivatives (for $P$-wave states) from the potential model calculations are $|R_{1S}(0)|^2=6.477~\textrm{GeV}^3$ and $|R'_{1P}(0)|^2=1.417 ~\textrm{GeV}^5$~\cite{bquarkmass2}, respectively. As for the renormalization scale $\mu_R$, we choose $2m_b$ to be its central value. We adopt the two-loop strong coupling constant to do our calculation, i.e.
\begin{eqnarray}
\alpha_s(\mu_{R})=\frac{4\pi}{\beta_0 L}-\frac{4\pi\beta_1 \ln L} {\beta_0^3 L^2}, \label{runningcouple1}
\end{eqnarray}
where $L=\ln(\mu^2_{R}/\Lambda^2_{\rm QCD})$, $\beta_0=11-\frac{2}{3}n_f$, and $\beta_1=\frac{2}{3}(153-19n_f)$ with the active flavor number $n_f=5$ and $\Lambda^{n_f=5}_{\rm QCD}=0.231$. Thus, we have $\alpha_s(2m_b)$=0.18. The fine-structure constant $\alpha=1/137$. Other input parameters are taken from the Particle Data Group~\cite{pdg}: $m_e=0.50 \times 10^{-3}$ GeV, $\Gamma_Z=2.4952$ GeV, $m_Z=91.1876$ GeV and ${\sin}^{2}{\theta _w}$=$0.2312$.

As a cross check of our calculation, when taking the same input parameters as those of Ref.\cite{ccY3}, we obtain the same tree-level and one-loop QCD cross sections at the $B$ factory, in which the channel $e^+e^- \to\gamma^*\to |H_{b\bar{b}}\rangle+\gamma$ provides dominant contribution and the channel $e^+e^- \to Z^0\to |H_{b\bar{b}}\rangle +\gamma$ is negligible.

\subsection{Total and differential cross sections for the bottomonium production at the super $Z$ factory}

\begin{table*}[htb]
\begin{tabular}{c| c c c c c c}
\hline
 &~$\eta_{b}$~&~$\Upsilon$~&~$h_b$~&~$\chi_{b0}$~&~$\chi_{b1}$~&~$\chi_{b2}$~\\
\hline
$\sigma_{\rm tree}$ (fb)~&~$1.27$~&~$51.41$~&~$7.34\times 10^{-2}$~&~$1.10\times 10^{-2}$~&~$7.19\times 10^{-2}$~&~$2.45\times 10^{-2}$~\\
\hline
$\sigma_{\rm QCD}$ (fb)~&~$1.25$~&~$36.51$~&~$4.14 \times 10^{-2}$~&~$1.15 \times 10^{-2}$~&~$6.83 \times 10^{-2}$~&~$1.17 \times 10^{-2}$~\\
\hline
\end{tabular}
\caption{Total cross sections for the bottomonium production via $e^+e^- \to\gamma^*/Z^0 \to |H_{b\bar{b}}\rangle +\gamma$ with or without one-loop QCD correction at the super $Z$ factory. We use the symbol `tree' to stand for the tree-level cross section and the symbol `QCD' to stand for the cross section with one-loop QCD correction. $\sqrt{s}=m_Z$, $m_b$=4.9 GeV and $\mu_R=2m_b$. }  \label{Totalcrosssection1}
\end{table*}

We put the total cross sections for the bottomonium production via $e^+e^- \to \gamma^*/Z^0\to |H_{b\bar{b}}\rangle+\gamma$ with or without one-loop QCD correction at the super $Z$ factory in Table \ref{Totalcrosssection1}, where the channels via a virtual photon or $Z^0$, including their cross-terms, have been taken into consideration simultaneously. For convenience, throughout the paper, we use the symbol `tree' to stand for the tree-level cross section and the symbol `QCD' to stand for the cross section with one-loop QCD correction. We have set $\mu_R=2m_b$ and $\sqrt{s}=m_Z$. By adding all $1S$-wave bottomonium states together, we obtain
\begin{eqnarray}
\sigma_{\rm tree}\left(e^+e^- \to |H_{b\bar{b}}\rangle(1S)+\gamma\right) &=& 52.68~\textrm{fb} \\
\sigma_{\rm QCD}\left(e^+e^- \to |H_{b\bar{b}}\rangle(1S)+\gamma\right)&=& 37.76~\textrm{fb}
\end{eqnarray}
And by adding all $1P$-wave bottomonium states together, we obtain
\begin{eqnarray}
\sigma_{\rm tree}\left(e^+e^- \to |H_{b\bar{b}}\rangle(1P) +\gamma\right) &=& 2.01~\textrm{fb} \\
\sigma_{\rm QCD}\left(e^+e^- \to |H_{b\bar{b}}\rangle(1P) +\gamma\right) &=& 1.33~\textrm{fb}
\end{eqnarray}
It is noted that the bottomonium cross sections are dominated by $\Upsilon$ production, which is due to the $t(u)$-channel diagrams and has already been pointed out by Ref.\cite{ccY6} at the tree level. By adding all the mentioned color-singlet bottomonium states together, we observe that the one-loop QCD correction for all bottomonium states is large and negative, $R\simeq-30\%$, where the ratio $R$ is defined as $(\sigma_{\rm QCD}-\sigma_{\rm tree})/\sigma_{\rm tree}$. This indicates the necessity and importance of the one-loop corrections. More specifically, for the cases of $\Upsilon$, $h_b$ and $\chi_{b2}$ production, the one-loop corrections are large, namely $R_{\Upsilon}=-29.0\%$, $R_{h_b}=-43.6\%$ and $R_{\chi_{b2}}=-52.2\%$. While for the cases of $\eta_b$, $\chi_{b0}$ and $\chi_{b1}$ production, the one-loop corrections are moderate, $R_{\eta_b}=-1.6\%$, $R_{\chi_{b0}}=4.5\%$ and $R_{\chi_{b1}}=-5.0\%$.

As for the usually analyzed $s$-channel only, for the bottomonium production via the channel $e^+e^- \to\gamma^*\to |H_{b\bar{b}}\rangle+\gamma$, it is found that only $\eta_b$ and $\chi_{bJ}$ can provide non-zero contributions at the super $Z$ factory. And even those non-zero cross sections are very small, i.e.
\begin{eqnarray}
\sigma_{\rm QCD}^{s,\gamma^*}(\eta_b+\gamma)   &=& 1.74 \times 10^{-3}~~\rm{fb}, \nonumber \\
\sigma_{\rm QCD}^{s,\gamma^*}(\chi_{b0}+\gamma)&=& 1.61 \times 10^{-5}~~\rm{fb}, \nonumber \\
\sigma_{\rm QCD}^{s,\gamma^*}(\chi_{b1}+\gamma)&=& 9.54 \times 10^{-5}~~\rm{fb}, \nonumber \\
\sigma_{\rm QCD}^{s,\gamma^*}(\chi_{b2}+\gamma)&=& 1.63 \times 10^{-5}~~\rm{fb}, \nonumber
\end{eqnarray}
where the superscripts $s$ and $\gamma^*$ indicate the cross section is for the process via the $s$-channel with virtual photon. Comparing with the total cross sections listed in Table \ref{Totalcrosssection1}, one can conclude that the total cross sections via $e^+e^- \to Z^0 \to |H_{b\bar{b}}\rangle+\gamma$ are dominant over those via $e^+e^- \to\gamma^*\to |H_{b\bar{b}}\rangle+\gamma$ by about three orders. This is caused by the $Z^0$ boson resonance effect, as is one of the advantage of the super $Z$ factory.

\begin{table}[htb]
\begin{tabular}{c |c c c}
\hline
 &~$s$-channel~&~$t(u)$-channel~ & ~Total~ \\
\hline
$\sigma_{\rm tree}$ (fb) ~&~$2.69$~&~$49.35$~&~$51.41$~ \\
\hline
$\sigma_{\rm QCD}$ (fb) ~&~$2.79$~&~$34.27$~&~$36.51$~ \\
\hline
\end{tabular}
\caption{Total cross sections for the $\Upsilon$ production via $e^+e^- \to \gamma^*/Z^0 \to \Upsilon+\gamma$ with or without one-loop QCD correction. We use the symbol 'tree' to stand for the tree-level cross section and the symbol `QCD' to stand for the cross section with one-loop QCD correction. $\sqrt{s}=m_Z$, $m_b$=4.9 GeV and $\mu_R=2m_b$. `Total' refers to the sum of $s$- and $t(u)$-channels including their cross terms. }
\label{Totalcrosssection3}
\end{table}

Moreover, it is found that for the $t(u)$-channels via $Z^0$ propagator, the $\eta_b$, $h_b$, $\chi_{b0}$ and $\chi_{b2}$ directly have zero contributions and those non-zero ones are also highly suppressed by the far off-shell $Z^0$ propagator in most of the final-state particles' kinematic regions, i.e.
\begin{eqnarray}
\sigma_{\rm tree}^{t,Z^0}(\Upsilon+\gamma)&=&1.30 \times 10^{-3}~~\rm{fb},\nonumber\\
\sigma_{\rm tree}^{t,Z^0}(\chi_{b1}+\gamma)&=&5.27 \times 10^{-5}~~\rm{fb}.\nonumber
\end{eqnarray}
For the $t(u)$-channels via $\gamma^*$ propagator, the conditions are similar. However, there is an important exception, it is noted that the $t(u)$-channel diagrams for $e^+e^- \to \gamma^* \to \Upsilon+\gamma$ provide quite large contribution. To show this point more clearly, we show the cross sections for $s$- and $t(u)$- channels of $e^+e^- \to \gamma^*/Z^0 \to \Upsilon+\gamma$ in Table \ref{Totalcrosssection3}. It indicates that the $t(u)$-channel are dominant for the $\Upsilon$ production, i.e. it increases the $s$-channel cross section by about eighteen times for the tree level and about twelve times for the one-loop QCD correction. The cross-terms between $s$- and $t(u)$-channels are sizable, whose magnitudes are about $20\%$ of $s$-channel cross sections. The `Total' cross sections in Table \ref{Totalcrosssection3} are the sum of $s$- and $t(u)$-channels including their cross terms.

\begin{figure}[htb]
\includegraphics[width=0.5\textwidth]{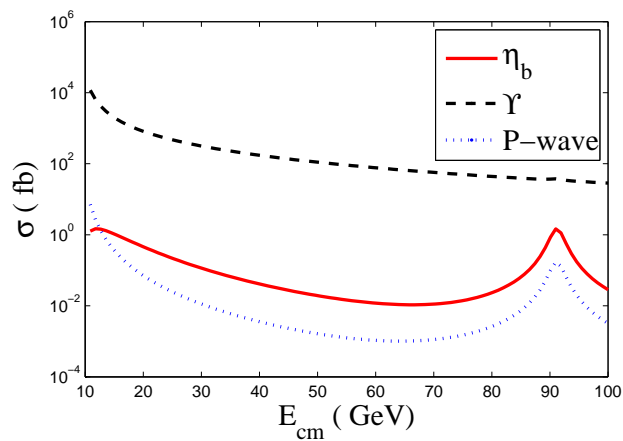}
\caption{Total cross sections with one-loop QCD correction to $e^+e^- \to Z^0/\gamma^* \to |H_{b\bar{b}}\rangle+\gamma$ versus the collision energy $E_{cm}$. $\mu_R=2m_b$. Here, all $P$-wave bottomonium states' cross sections have similar behaviors and have been summed up.}  \label{ecmrunning}
\end{figure}

To show how the total cross sections change with the $e^+e^-$ collision energy $E_{cm}(=\sqrt{s})$, we present the total cross sections versus $E_{cm}$ with one-loop QCD correction in Fig.(\ref{ecmrunning}). As for $\eta_b$ and $P$-wave bottomonium states production, in most of the $e^+e^-$ collision energies, they are dominated by the $s$-channel diagrams of $e^+e^- \to Z^0 \to |H_{b\bar{b}}\rangle+\gamma$, and the $Z^0$ boson resonance effect are significant at $E_{cm}=m_Z$. While for the $\Upsilon$ production, its $Z^0$ boson resonance effect is smeared by the large contributions from the $t(u)$-channel diagrams of $e^+e^- \to \gamma^* \to \Upsilon+\gamma$.

\begin{center}
\begin{table}[htb]
\begin{tabular}{c| c c c}
\hline
 &~$|\cos\theta|\leq 1$~&~$|\cos\theta| \leq 0.9$~&~$|\cos\theta|\leq 0.8$~\\
\hline
~~$\sigma_{\eta_b+\gamma}$~~&~$1.25$~&~$1.07$~&~$9.09 \times 10^{-1}$~\\
$\sigma_{\Upsilon+\gamma}$~&~$36.51$~&~$4.95$~&~$3.69$~\\
$\sigma_{h_b+\gamma}$~&~$4.14\times 10^{-2}$~&~$3.55\times 10^{-2}$~&~$3.02\times 10^{-2}$~\\
$\sigma_{\chi_{b0}+\gamma}$~&~$1.15\times 10^{-2}$~&~$9.84\times 10^{-3}$~&~$8.36\times 10^{-3}$~\\
$\sigma_{\chi_{b1}+\gamma}$~&~$6.83\times 10^{-2}$~&~$5.86\times 10^{-2}$~&~$4.99\times 10^{-2}$~\\
$\sigma_{\chi_{b2}+\gamma}$~&~$1.17\times 10^{-2}$~&~$1.00\times 10^{-2}$~&~$8.55\times 10^{-3}$~\\
\hline
\end{tabular}
\caption{Total cross sections with one-loop QCD correction (in fb) for the bottomonium production via $e^+e^- \to\gamma^*/Z^0\to |H_{b\bar{b}}\rangle+\gamma$ with different cuts on $|\cos\theta|$. $\sqrt s = m_Z$, $m_b=4.9$ GeV and $\mu_R=2m_b$.} \label{coscut}
\end{table}
\end{center}

\begin{figure}[htb]
\includegraphics[width=0.5\textwidth]{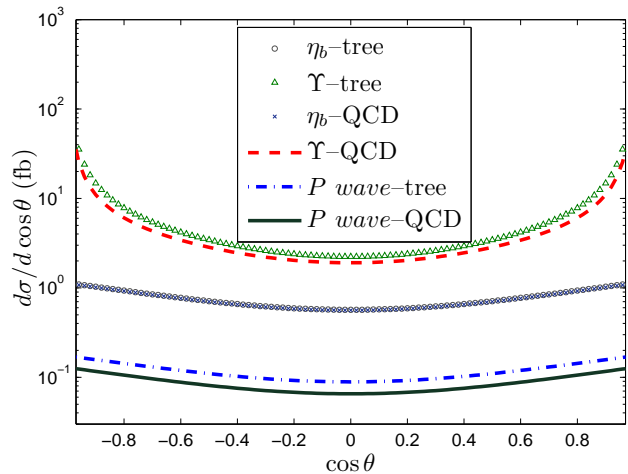}
\caption{Differential cross sections for $e^+e^- \to\gamma^*/Z^0\to |H_{b\bar{b}}\rangle +\gamma$ versus $\cos\theta$. Here, all $P$-wave bottomonium states' cross sections have similar behaviors and have been summed up. We use the symbol `tree' to stand for the tree-level differential cross section and the symbol `QCD' to stand for the differential cross section with one-loop QCD correction. For the $\eta_b$ production, its tree-level and one-loop curves are almost coincide with each other. $\sqrt s=m_Z$, $m_b=4.9$ GeV and $\mu_R=2m_b$. } \label{cosdistribution}
\end{figure}

Experimentally, no detector can cover all the kinematics of the events, thus, only some of them can be observed. For example, the bottomonium events with $|\cos\theta| \to 1$ can not be detected, since under such condition, the produced bottomonium moves very close to the beam direction. Here $\theta$ stands for the angle between the three-vector momentums of the bottomonium and the electron. Considering the detector's abilities and in order to offer experimental references, we try various cuts on $\cos\theta$, the results are listed in Table \ref{coscut}. As for $\eta_b$ and $P$-wave bottomonium production, the changes of the cross section over the $|\cos\theta|$ cut are moderate: setting $|\cos\theta|\leq0.9$, the cross sections are changed by $\sim14\%$; setting $|\cos\theta|\leq0.8$, the cross sections are further changed by $\sim15\%$. While for $\Upsilon$ production, when setting $|\cos\theta|\leq0.9$, the cross section is changed by $86\%$; when setting $|\cos\theta|\leq0.8$, the cross section is further changed by $25\%$. Such great differences can be explained by comparing with their differential cross sections, which are presented in Fig.(\ref{cosdistribution}).

At the super $Z$ factory, with the high luminosity ${\cal L}=10^{36} {\rm cm}^{-2} {\rm s}^{-1}$ (or $\simeq 10^{4}~{\rm fb}^{-1}$ in one operation year), large amount of bottomonium can be generated. Thus, the super $Z$ factory could be a useful platform for studying bottomonium properties, especially for $\eta_b$ and $\Upsilon$. More explicitly, without applying any kinematical cut, we obtain
\begin{eqnarray}
&&N_{\eta_b}=1.25 \times 10^{4},~~~N_{\Upsilon}=3.65 \times 10^{5}, \nonumber\\
&&N_{h_b}=4.14 \times 10^{2},~~~N_{\chi_{b0}}=1.15 \times 10^{2}, \nonumber\\
&&N_{\chi_{b1}}=6.83 \times 10^{2},~~N_{\chi_{b2}}=1.17 \times 10^{2}.
\end{eqnarray}
If setting $|\cos\theta|\leq 0.9$, we obtain
\begin{eqnarray}
&&N_{\eta_b}=1.07 \times 10^{4},~~~N_{\Upsilon}= 4.95 \times 10^{4}, \nonumber\\
&&N_{h_b}=3.55 \times 10^{2},~~~N_{\chi_{b0}}=9.84 \times 10^{1}, \nonumber\\
&&N_{\chi_{b1}}=5.86 \times 10^{2},~~N_{\chi_{b2}}=1.00 \times 10^{2} .
\end{eqnarray}
If setting $|\cos\theta|\leq 0.8$, we obtain
\begin{eqnarray}
&&N_{\eta_b}=9.09 \times 10^{3},~~~N_{\Upsilon}=3.69 \times 10^{4}, \nonumber\\
&&N_{h_b}=3.02 \times 10^{2},~~~N_{\chi_{b0}}=8.36 \times 10^{1}, \nonumber\\
&&N_{\chi_{b1}}=4.99 \times 10^{2},~~N_{\chi_{b2}}=8.55 \times 10^{1} .
\end{eqnarray}

\subsection{Theoretical uncertainties}

\begin{table}[htb]
\begin{center}
\begin{tabular}{c| c c c}
\hline
$E_{cm}$~&~$m_Z-0.5$ GeV ~& ~$m_Z$~&~$m_Z+0.5$ GeV~\\
\hline
$\sigma_{\eta_b+\gamma}$~&~$1.08$~&~$1.25$~&~$1.08$~\\
$\sigma_{\Upsilon+\gamma}$~&~$36.53$~&~$36.51$~&~$35.81$~\\
$\sigma_{h_b+\gamma}$~&~$3.57\times 10^{-2}$~&~$4.14 \times 10^{-2}$~&~$3.56\times 10^{-2}$~\\
$\sigma_{\chi_{b0}+\gamma}$~&~$9.90\times 10^{-3}$~&~$1.15 \times 10^{-2}$~&~$9.90\times 10^{-3}$~\\
$\sigma_{\chi_{b1}+\gamma}$~&~$5.88\times 10^{-2}$~&~$6.83 \times 10^{-2}$~&~$5.88\times 10^{-2}$~\\
$\sigma_{\chi_{b2}+\gamma}$~&~$1.01\times 10^{-2}$~&~$1.17 \times 10^{-2}$~&~$1.01\times 10^{-2}$~\\
\hline
\end{tabular}
\caption{Total cross sections with one-loop QCD correction (in fb) for the bottomonium production via $e^+e^- \to\gamma^*/Z^0\to |H_{b\bar{b}}\rangle+\gamma$ under three $e^+e^-$ collision energies at the super $Z$ factory. $m_b=4.9$ GeV and $\mu_R=2m_b$.} \label{ecmuncertainty}
\end{center}
\end{table}

It is noted that the productions are likely to be affected by the resummation of initial state electromagnetic radiation, which may shift (lower) the effective $e^+ e^-$ collision energy slightly away from the $Z^0$ resonance peak to a certain degree, cf. Refs.\cite{Zini1,Zini2,Zini3}. Because of the importance of the $Z^0$ boson resonance effect to the present production, such a shift may heavily affect the productions. It is then helpful to know how the total cross sections are affected by varying the collision energy $E_{cm}$ away from $Z^0$ peak. For the purpose, we calculate the total cross sections by varying $E_{cm}=m_Z\pm 0.5$ GeV. The results are presented in Table \ref{ecmuncertainty}. Consistent with the above observations from Fig.(\ref{ecmrunning}), i.e. the $Z^0$ boson resonance effect is sizable for the $\eta_b$ and $P$-wave bottomonium production, their total cross sections drop down by $\sim14\%$ from their central values when varying $E_{cm}$ by merely $0.5$ GeV. While under the same energy changes, the total cross section for the $\Upsilon$ production changes only by less than $2\%$.

\begin{center}
\begin{table}[htb]
\begin{tabular}{c| c c c}
\hline
~~$m_b$~~ &~$4.8$ GeV~&~$4.9$~GeV~&~$5.0$ GeV~\\
\hline
$\sigma_{\eta_b+\gamma}$~&~$1.28$~&~$1.25$~&~$1.22$~\\
$\sigma_{\Upsilon+\gamma}$~&~$38.66$~&~$36.51$~&~$34.53$~\\
$\sigma_{h_b+\gamma}$~&~$4.38\times 10^{-2}$~&~$4.14 \times 10^{-2}$~&~$3.91\times 10^{-2}$~\\
$\sigma_{\chi_{b0}+\gamma}$~&~$1.22\times 10^{-2}$~&~$1.15 \times 10^{-2}$~&~$1.08\times 10^{-2}$~\\
$\sigma_{\chi_{b1}+\gamma}$~&~$7.27\times 10^{-2}$~&~$6.83 \times 10^{-2}$~&~$6.41\times 10^{-2}$~\\
$\sigma_{\chi_{b2}+\gamma}$~&~$1.23\times 10^{-2}$~&~$1.17 \times 10^{-2}$~&~$1.11\times 10^{-2}$~\\
\hline
\end{tabular}
\caption{Total cross sections with one-loop QCD correction (in fb) for the bottomonium production channel $e^+e^- \to \gamma^*/Z^0\to |H_{b\bar{b}}\rangle+\gamma$ for $m_b=(4.9\pm0.1)$ GeV. $\sqrt s=m_Z$ and $\mu_R=2m_b$. } \label{massuncertainty}
\end{table}
\end{center}

Different choice of the effective $b$ quark mass shall also lead to sizable changes to the total cross section. We put the results for $m_b=(4.9\pm0.1)$ GeV in Table \ref{massuncertainty}. It is found that when the $b$ quark mass varies within such region, the total cross sections change by about $2\%-7\%$ from their central values.

\begin{center}
\begin{table}[htb]
\begin{tabular}{c| c c c}
\hline
 &~$\mu_R=m_b$~&~$\mu_R=2m_b$~&~$\mu_R={\sqrt s}/{2}$~\\
\hline
~~$\sigma_{\eta_b+\gamma}$~~&~$1.24$~&~$1.25$~&~$1.26$~\\
$\sigma_{\Upsilon+\gamma}$~&~$33.53$~&~$36.51$~&~$40.46$~\\
$\sigma_{h_b+\gamma}$~&~$3.50\times 10^{-2}$~&~$4.14\times 10^{-2}$~&~$4.99\times 10^{-2}$~\\
$\sigma_{\chi_{b0}+\gamma}$~&~$1.16\times 10^{-2}$~&~$1.15\times 10^{-2}$~&~$1.14\times 10^{-2}$~\\
$\sigma_{\chi_{b1}+\gamma}$~&~$6.75\times 10^{-2}$~&~$6.83\times 10^{-2}$~&~$6.92\times 10^{-2}$~\\
$\sigma_{\chi_{b2}+\gamma}$~&~$9.11\times 10^{-3}$~&~$1.17\times 10^{-2}$~&~$1.51\times 10^{-2}$~\\
\hline
\end{tabular}
\caption{Conventional scale uncertainties for the total cross sections (in fb) for $e^+e^- \to\gamma^*/Z^0 \to |H_{b\bar{b}}\rangle+\gamma$ with one-loop QCD correction by adopting three typical renormalization scales. $\sqrt{s}=m_Z$ and $m_b$=4.9 GeV. } \label{Totalcrosssection4}
\end{table}
\end{center}

\begin{figure}[htb]
\includegraphics[width=0.5\textwidth]{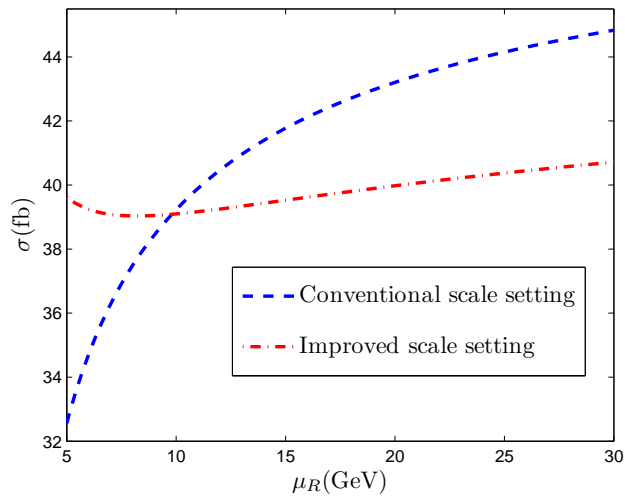}
\caption{Total cross sections for the bottomonium production via $e^+e^- \to \gamma^*/Z^0\to |H_{b\bar{b}}\rangle+\gamma$ with one-loop QCD correction versus the scale $\mu_R$, where all bottomonium states have been summed up. As a comparison, we present the cross sections under both the conventional scale setting and the improved scale setting, respectively. $\sqrt s=m_Z$ and $\mu_R=2m_b$. } \label{runningcopling}
\end{figure}

As an estimation of the physical observable, one can take a typical momentum flow of the process as the scale and vary it within certain range, or directly take several typical scales, to study the renormalization scale uncertainty. At present, we take three frequently used scales, $m_b$, $2m_b$ and $\sqrt s/2$, to study the conventional scale uncertainty. The results are presented in Table \ref{Totalcrosssection4}. At the one-loop level, the renormalization scale uncertainty can be up to $30\%$ (for the case of $\chi_{b2}$).

As has been suggested in Ref.\cite{PMC2}, the principle of maximum conformality (PMC) provides a systematic way to eliminate the renormalization scale ambiguity. It however depends on how well we know the $\beta$-terms in pQCD series. Those $\beta$-terms rightly determine the running coupling behavior via the renormalization group equation and fix the renormalization scale to a certain degree~\cite{PMC2}. Strictly, we need to finish a two-loop QCD calculation for the present considered bottomonium production process such that we can find out the $\{\beta_i\}$-terms and set the PMC scale for the one-loop terms. Such a two-loop calculation is not available at the present. As a compromise, we can use an improved way suggested in Ref.\cite{PMC1} to reestimate the renormalization scale uncertainty, in which the one-higher order $\beta$-terms are directly determined from the renormalization group equation. A comparison with those two approaches is presented in Fig.(\ref{runningcopling}). Fig.(\ref{runningcopling}) shows that a more reliable renormalization scale uncertainty can indeed be achieved by using the improved scale setting approach.

As has already been observed, the $t(u)$-channel diagrams provide a dominant contribution to $\Upsilon$ production. Such large $t(u)$-channel contributions for the $\Upsilon$ production are reasonable, since there are large contributions from the kinematic region when the photons are almost collinear to the incident electron line. Usually, the electron mass is neglected in doing the analytic calculations. For the present case, to get a steady and reliable estimation, we keep all those terms that are proportional to $m_e$ and take $m_e=0.50\times10^{-3}$ GeV to do our calculation~\cite{pdg} \footnote{A steady cross section can indeed be obtained, i.e. the cross sections at both the tree-level and the one-loop QCD correction shall only be changed by less than $\pm 2\%$ even by varying $m_e\in[0.40,0.60]\times10^{-3}$ GeV.}.

\subsection{An estimation of physical background}

\begin{figure}[tb]
\includegraphics[width=0.5\textwidth]{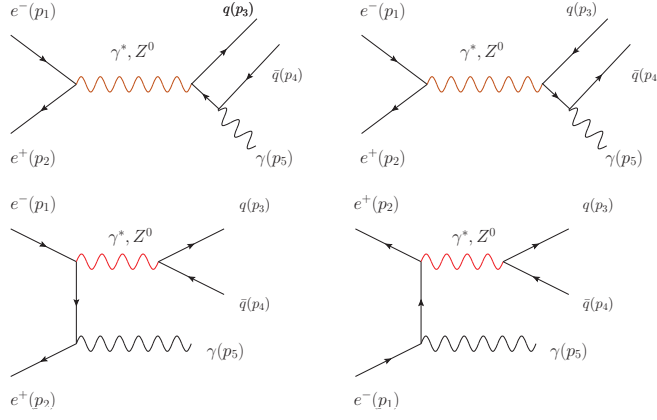}
\caption{Tree-level Feynman diagrams for $e^+e^- \to\gamma^*/Z^0\to  q+\bar{q}+\gamma$, where $q$ stands for $u$, $d$, $s$, $c$ or $b$ quark, respectively.}\label{BFFD}
\end{figure}

Experimentally, one may observe the process $e^+e^- \to |H_{b\bar{b}}\rangle +\gamma$ by analyzing the energy spectrum of the radiated photons. The dominant background subprocesses are $e^+e^- \to q+\bar{q} +\gamma$~\cite{ccY5}, the Feynman diagrams are shown in Fig.(\ref{BFFD}), where $u$, $d$, $s$, $c$ or $b$ quark, respectively. To estimate the background, we take: $m_u=2.3$ MeV, $m_d=4.8$ MeV, $m_s=95$ MeV and $m_e=0.5\times10^{-3}$ GeV~\cite{pdg}. We restrict the energy range of the radiated photons as $45.067 \pm 0.005 \textrm{GeV}$, where the central value is determined by the process $e^+e^- \to |H_{b\bar{b}}\rangle +\gamma$, since its radiated photon rightly has a fixed energy $E_\gamma= {(s-m^2_{H_{b\bar{b}}})} / {(2\sqrt{s})}=45.067$ GeV by taking $m_{H_{b\bar{b}}}=2m_b$ and $\sqrt{s}=m_Z$.

The total cross sections for the background process $e^+e^- \to q+\bar{q} +\gamma$ are $\sigma_{BG}\simeq130$ fb for $|\cos\theta|\leq1$, $\sigma_{BG}\simeq59$ fb for $|\cos\theta|\leq0.9$ and $\sigma_{BG}\simeq50$ fb for $|\cos\theta|\leq0.8$. These background cross sections are greater than the wanted signal cross sections shown in Tables \ref{Totalcrosssection1} and \ref{coscut}. However, the high luminosity super $Z$ factory might still allow one to measure the bottomonium production associated with a photon. As a rough estimation, we can compute the signal significance $S(H)$ for the bottomonium production plus one photon, which is defined as $S(H)=N_H/\sqrt{N_{BG}}$. At the super $Z$ factory, taking the luminosity ${\cal L}=10^{36}{\rm cm}^{-2}{\rm s}^{-1}$, then, if setting $|\cos\theta|\leq 1$, we have $S(\eta_b)=11$, $S(\Upsilon)=320$ and $S(1P)=1.2$, where $S(1P)$ stands for the sum of the mentioned $P$-wave bottomonium states; if setting $|\cos\theta|\leq 0.9$, we have $S(\eta_b)=14$, $S(\Upsilon)=64$ and $S(1P)=1.5$, where $S(1P)$ stands for the sum of the mentioned $P$-wave bottomonium states.

\section{Summary}

At the super $Z$ factory, we can obtain a more reliable pQCD estimation for the bottomonium properties and a more confidential test of NRQCD factorization theorem than the $B$ factory. In the present paper, we have studied the bottomonium production via $e^{+}e^{-} \to\gamma^*/Z^0 \to |H_{b\bar{b}}\rangle + \gamma$ with one-loop QCD correction at the super $Z$ factory. By adding all the mentioned color-singlet bottomonium states together, we observe that the one-loop QCD correction for all bottomonium states is large and negative, $R=(\sigma_{\rm NLO}-\sigma_{\rm LO})/\sigma_{\rm LO}\simeq-30\%$. This indicates the necessity and importance of the one-loop corrections. Especially, for the cases of $\Upsilon$, $h_b$ and $\chi_{b2}$ production, whose one-loop corrections are large, i.e. $R_{\Upsilon}=-29.0\%$, $R_{h_b}=-43.6\%$ and $R_{\chi_{b2}}=-52.2\%$.

\begin{figure}[htb]
\includegraphics[width=0.5\textwidth]{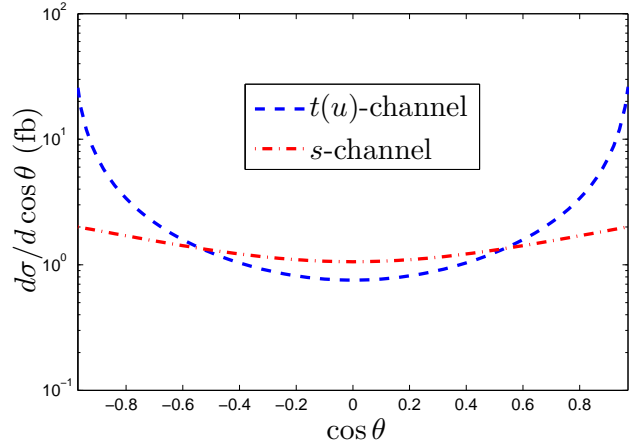}
\caption{A comparison of $s$- and $t(u)$-channel differential cross sections with one-loop QCD correction for $\Upsilon$ production via $e^{+}e^{-} \to\gamma^*/Z^0 \to \Upsilon + \gamma$. $E_{cm}=m_Z$, $m_b=4.9$ GeV and $\mu_R=2m_b$. } \label{cosdistribution1}
\end{figure}

In low $e^+ e^-$ collision energy, e.g. at the $B$ factory, the channel $e^+e^- \to\gamma^*\to |H_{b\bar{b}}\rangle+\gamma$ provides dominant contribution and the channel $e^+e^- \to Z^0\to |H_{b\bar{b}}\rangle +\gamma$ is negligible. Because the bottomonium mass is close to the threshold of the $B$ factories as Belle and BABAR, the emitted photons could be soft. Thus, the pQCD estimations on the bottomonium production at the $B$ factories is questionable.

On the other hand, a more confidential estimation can be achieved at the super $Z$ factory. Due to $Z^0$ boson resonance effect, the process $e^+e^- \to Z^0\to |H_{b\bar{b}}\rangle +\gamma$ shall dominant over the process $e^+e^- \to\gamma^*\to |H_{b\bar{b}}\rangle+\gamma$ for the bottomonium states such as $\eta_b$, $h_b$ and $\chi_{bJ}$. In fact, such a $Z^0$ boson resonance effect is very important for measuring these states, a slight change of $E_{cm}$ from $m_Z$ by merely $0.5$ GeV shall reduce their total cross sections by $\sim 14\%$ from their central values. The only exception is the $\Upsilon$ production, whose $t(u)$-channel diagrams provide dominant contribution to the total cross section as shown by Table \ref{Totalcrosssection3}. More specifically, Fig.(\ref{cosdistribution1}) shows the relative importance of the $s$-channel and $t(u)$-channel contributions. It shows that the $t(u)$-channel can provide significant contributions to the $\Upsilon$ production at the whole kinematic region.

At the super $Z$ factory with a high luminosity up to ${\cal L}=10^{36}{\rm cm}^{-2}{\rm s}^{-1}$, the bottomonium production events are sizable. By summing all bottomonium states together, in one operation year, we shall have $3.8\times10^{5}$ bottomonium events for $|\cos\theta|\leq1$, $6.1\times10^{4}$ bottomonium events for $|\cos\theta|\leq0.9$ and $4.7\times10^{4}$ bottomonium events for $|\cos\theta|\leq0.8$. Thus, the high luminosity super $Z$ factory may allow one to measure the bottomonium properties, even though one needs to deal with the background processes carefully.

\begin{figure}[tb]
\includegraphics[width=0.5\textwidth]{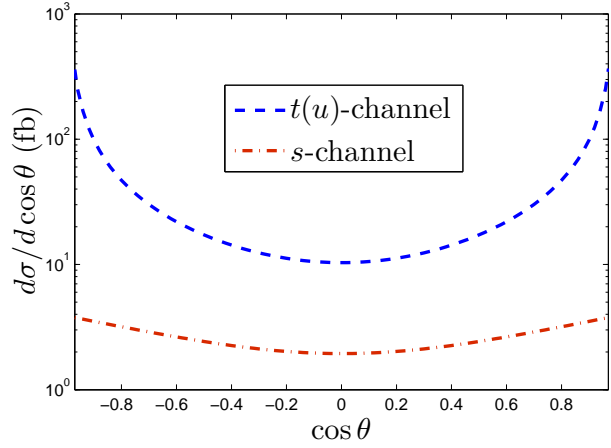}
\caption{A comparison of $s$- and $t(u)$-channel differential cross sections with one-loop QCD correction for $J/\psi$ production via $e^{+}e^{-} \to\gamma^*/Z^0 \to J/\psi + \gamma$. $E_{cm}=m_Z$, $m_c=1.5$ GeV and $\mu_R=2m_c$. } \label{cosdistribution2}
\end{figure}

As a final remark, we put a comparison of $s$- and $t(u)$-channel distributions versus $\cos\theta$ for the charmonium production in Fig.(\ref{cosdistribution2}). The charmonium is produced via the channel $e^{+}e^{-} \to\gamma^*/Z^0 \to J/\psi + \gamma$. It shows that the $t(u)$-channel diagrams are also important for producing charmonium events, since their distributions are at least five times bigger than the $s$-channel distribution at each $\cos\theta$-point in the final particles' whole kinematic region.

\hspace{2cm}

{\bf Acknowledgement:} The authors are grateful for the anonymous referee's comments and suggestions that substantially improve the paper. This work was supported in part by the Fundamental Research Funds for the Central Universities under Grant No.CQDXWL-2012-Z002, the Program for New Century Excellent Talents in University under Grant No.NCET-10-0882, and the Natural Science Foundation of China under Grant No.11275280.

\appendix

\section{Analytical results for the one-loop integrals}

For simplicity, we take ${\cal I}^{(n)}$ to denote the master integrals, where $n=1$, $2$ and $3$ for one-point, two-point and three-point scalar integrals, respectively.
\begin{widetext}
\begin{eqnarray}
{\cal I}^{(1)}&=& N \int\frac{\textrm{d}^Dl}{\left(l\pm\frac{q_1}{2} \right)^2-m_b^2}=N \int\frac{\textrm{d}^Dl}
{\left(l\pm\frac{p_3}{2}\pm p_4\right)^2-m_b^2} = N_1 m_b^2 \left[\ln\left(\frac{\mu^2}{m_b^2} \right)+\frac{1}{\epsilon }-\gamma_{E} +1 \right]\;, \nonumber \\
{\cal I}^{(2)}_{1}&=& N \int\frac{\textrm{d}^Dl}{l^2 \left[\left(l\pm\frac{p_3}{2}\pm p_4\right)^2-m_b^2 \right]} = N_1 \left[\ln \left(\frac{\mu^2}{m_b^2}\right)-\frac{2r-2 }{2 r-1}\ln (2r-2)+\frac{1}{\epsilon }-\gamma_{E} +2 \right] \;,\nonumber \\
{\cal I}^{(2)}_{2}&=& N \int\frac{\textrm{d}^Dl}{\left(l^2-m_b^2\right) \left[\left(l-p_3- p_4\right)^2-m_b^2 \right]} = N_1 \left[ \ln \left(\frac{\mu^2}{s}\right)+(a-b)\ln\left(\frac{b}{a}\right)-\ln(ab)+ \frac{1}{\epsilon }-\gamma_E +2 \right] \;\nonumber \\
{\cal I}^{(3)}_{1} &=& N \int\frac{\textrm{d}^Dl}{l^2 \left[\left(l+\frac{p_3}{2}\right)^2-m_b^2 \right]\left[\left(l - \frac{p_3}{2} - p_4\right)^2-m_b^2 \right]} = \frac{N_1}{s-4 m_b^2}\left[2\textrm{Li}_2\left(\frac{1}{2r-1}\right)+\ln ^2(2r-1)-\frac{\pi ^2}{3}\right]\;,\nonumber \\
{\cal I}^{(3)}_{2} &=& N \int\frac{\textrm{d}^Dl}{l^2 \left[\left(l - \frac{p_3}{2}\right)^2-m_b^2 \right]\left[\left(l - \frac{p_3}{2} - p_4\right)^2-m_b^2 \right]} \nonumber \\
&=&\frac{N_1}{s-4 m_b^2}\left[2 \textrm{Li}_2 (a) +2 \textrm{Li}_2(b)-2 \textrm{Li}_2\left(\frac{1}{2 r-1}\right)+\ln^2(a)+\ln^2(b)-\ln ^2(2r-1)\right]\;,\nonumber \\
{\cal I}^{(3)}_{3} &=& N \int\frac{\textrm{d}^Dl}{\left(l ^2-m_b^2 \right) \left[\left(l - p_3\right)^2-m_b^2 \right] \left[\left(l - p_3 - p_4\right)^2-m_b^2 \right]} \nonumber\\
&=&\frac{N_1}{s-4 m_b^2}\left[\textrm{Li}_2(a)+ \textrm{Li}_2(b) + \frac{\ln^2(a)}{2} + \frac{\ln^2(b)}{2} - \frac{\pi ^2}{6} \right]\;,
\end{eqnarray}
where
\begin{eqnarray}
N=\frac{\mu^{2\epsilon}\Gamma(\epsilon)}{(4\pi)^{2-\epsilon}},\; N_1=\frac{1}{(4\pi)^{2-\epsilon}},\;
r = \frac{s}{4m_b^2},\;
a = \frac{1}{2}\left(1+\sqrt{\frac{r-1}{r}}\right),\;
b = \frac{1}{2}\left(1-\sqrt{\frac{r-1}{r}}\right).\;
\end{eqnarray}
\end{widetext}

\end{document}